

\input phyzzx
\tolerance=10000

\def\sqr#1#2{{\vcenter{\hrule height.#2pt
      \hbox{\vrule width.#2pt height#1pt \kern#1pt
        \vrule width.#2pt}
      \hrule height.#2pt}}}

\def\gtorder{\mathrel{\raise.3ex\hbox{$>$}\mkern-14mu
             \lower0.6ex\hbox{$\sim$}}}
\def\ltorder{\mathrel{\raise.3ex\hbox{$<$}\mkern-14mu
             \lower0.6ex\hbox{\sim$}}}
\def\dalemb#1#2{{\vbox{\hrule height .#2pt
        \hbox{\vrule width.#2pt height#1pt \kern#1pt
                \vrule width.#2pt}
        \hrule height.#2pt}}}


\date = {July 1992.}
 \pubtype={QMW-PH-92-13.}
\titlepage
\title {{\bf THE SEARCH FOR ZOO-PERPARTICLES}\footnote\spadesuit
                                {This manuscript is based on a presentation
                                 given at the {\it Physique en Herbe 92}
                                 Conference held in Marseille, 6-10 July
                                 1992.}}

\author { Jose-Luis V\'azquez-Bello\footnote\S
                    {Work partially supported by private funds.}}

\address {Physics Department, Queen Mary and Westfield College,\break
             Mile End Road, London E1 4NS,\break
                 UNITED KINGDOM.}


\abstract
{This paper reviews the covariant formalism of N=1, D=10 classical
superparticle models. It discusses the local invariances of a number
of superparticle actions and highlights the problem of finding a covariant
quantization scenario. Covariant quantization has proved problematic,
but it has motivated in seeking alternative approaches that avoids
those found in earlier models. It also shows new covariant superparticle
theories formulated in extended spaces that preserve certain canonical
form in phase-space, and easy to quantize by using the Batalin-Vilkovisky
procedure, as the gauge algebra of their constraints only closes on-shell.
The mechanics actions describe particles moving in a superspace
consisting of the
usual $N=1$ superspace, together with an extra spinor or vector
coordinate.
A light-cone analysis shows that all these new superparticle models reproduce
the physical spectrum of the N=1 super-Yang-Mills theory.}

\endpage
\pagenumber=1
%
%
\def\psh{\rlap{/}{p}}
%
%

%
%
\def\np{Nucl. Phys.}
\def\pl{Phys. Lett.}
\def\pr{Phys. Rev.}

\def\ijmp{Intern. J. Mod. Phys.}
\def\mpl{Mod. Phys. Lett.}
\def\nc{Nuovo Cim.}

\def\cqg {Class. Quantum Grav.}
\def\jmp {J. Math. Phys.}

\def\ptps{Prog. Theor. Phys. Suppl.}

\def\rmf{Rev. Mex. de Fis.}
\def\cjm{Canad. J. Math.}
%
%
\def\half{{\textstyle {1\over 2}}}
\def\third{{\textstyle {1\over 3}}}

\def\inttau {\int\!d\tau }

\def\gammu {\gamma^\mu}
\def\pmu {p_\mu}
\def\xdot {\dot x}
\def\xmu{x^\mu}
\def\xplus{x^+}
\def\xmin{x^-}
\def\xpm{x^{\pm}}
\def\grad {\partial}
\def\thetaA {\theta_A}
\def\thetaB {\theta_B}
\def\kappaA {\kappa^A}

\def\psiA {\psi^A}
%
%

\def\third{{\textstyle {1\over 3}}}

\def\half {\textstyle {1 \over 2 }}
\def\seventh {\textstyle {1 \over 7 }}

\def\inttau {\int\! d\tau}
\def\xmu {x^\mu}
\def\xplus {x^{+}}

\def\phiA {\phi^A}

\def\hatphiA {{\hat\phi}_A}

\def\thetaA {\theta_A}
\def\thetaB {\theta_B}

\def\psiA {\psi^A}

\def\grad {\partial}

\def\pmu {p^\mu}

\def\gammu {\gamma^\mu}
\def\Gammu {\Gamma^\mu}

\def\Lamnrs {\Lambda_{\mu\nu\rho\sigma}}

\def\Gamnrs {\Gamma^{\mu\nu\rho\sigma}}

\def\rhomunu {\rho^{(\mu\nu )}}

\def\hatphi {{\hat\phi}}
\def\phimu {\phi^\mu}
\def\phinu {\phi^\nu}

\def\dA {d^A}
\def\dB {d^B}

\def\Upsimupha {{\Upsilon^\mu}_A }
\def\Cmupha {{C_\mu}^A}

%
%
\REF\kugo{T. Kugo and I. Ojima, \ptps\ {\bf 66} (1979) 1.}
\REF\popov{L. D. Fadeev and V. N. Popov, \pl\ {\bf B25} (1967) 29.}
\REF\BV{I. A. Batalin and G. A. Vilkovisky, \pl\ {\bf B102} (1981) 27;
                                            \pr\ {\bf D28} (1983) 2567.}
\REF\book{W. Siegel, {\it Introduction to String Field Theory}, ( World
                Scientific, 1988).}
\REF\casal{R. Casalbuoni, \pl\ {\bf B62} (1976) 49;
                          \nc\ {\bf A33} (1976) 389;
           P. G. O. Freund, unpublished, as quoted in A. Ferber, \np\
                               {\bf B132} (1978) 55.}
\REF\brink{L. Brink and J. H. Schwarz, \pl\ {\bf B100} (1981) 310;
           L. Brink and M. B. Green, \pl\ {\bf B106} (1981) 393.}
\REF\polyakov{L. Brink, {\it Covariant Formulation of the Superparticle
               and Superstring}, in Physics and Mathematics of Strings, eds.
               L. Brink, D. Friedan and A. M. Polyakov,
              ( World Scientific -1990 ).}
\REF\romans{L. J. Romans, \np\ {\bf B281} (1987) 639.}
\REF\begs{I. Bengtsson, \pr\ {\bf D39} (1989) 1158.}
\REF\huqone{M. Huq, {\it On the Covariant Quantization of Massive
                     Superparticle}, \np\ {\bf B315} (1989) 249.}
\REF\delius{F. Bastianelli, G. W. Delius and E. Laenen,
                                \pl\ {\bf B229} (1989) 223;
            J. M. L. Fisch and M. Henneaux, {\it A Note on the Covariant
                    BRST Quantization of the Superparticle}, Preprint
                    ULB-TH2/89-04.}
\REF\siegone{W. Siegel, \cqg\ {\bf 2} (1985) L95.}
\REF\siegel{W. Siegel, \np\ {\bf B263} (1986) 93.}
\REF\kallosh{E. A. Bergshoeff and R. E. Kallosh, \np\ {\bf B333} (1990) 605;
                                                \pl\ {\bf B240} (1990) 105;
             E. A. Bergshoeff, R. E. Kallosh and A. Van Proeyen,
                                                \pl\ {\bf B251} (1990) 128;
             R. E. Kallosh, \pl\ {\bf B251} (1990) 134.}
\REF\markov{I. A. Batalin, R. E. Kallosh and A. Van Proeyen, {\it Symmetries
              of Superparticle and Superstring Actions}, in Quantum
              Gravity, eds. M. A. Markov, V. A. Berezin and F. P. Frolov,
              (World Scientific, 1988).}
\REF\twisted{M. B. Green and C. M. Hull, {\it Quantum Mechanics of a Twisted
                             Superparticle }, \np\ {\bf B344} (1990) 115.}
\REF\linds{U. Lindstr\"om, M. Ro\v cek, W. Siegel, P. Van Nieuwenhuizen and
           A. E. Van de Ven, \jmp\ {\bf 31} (1990) 1761.}
\REF\huq{M. Huq, {\it Covariant Quantization of Broken Twisted N=2
                  Superparticle}, Preprint DOE-ER40200-257, June-1991.}
\REF\unoilk{F. E$\beta$ler, E. Laenen, W. Siegel and J. P. Yamron,
                       {\it BRST Operator for the first-ilk Superparticle},
                       Preprint ITP-SB-90-76.}
\REF\covilk{F. E$\beta$ler, M. Hatsuda, E. Laenen, W. Siegel, J. P. Yamron,
            T. Kimura and A. Mikovi\'c,
                 {\it Covariant Quantization of the first-ilk
                         Superparticle}, Preprint ITP-SB-90-77.}
\REF\belli{J. L. V\'azquez-Bello, \ijmp\ {\bf A19} (1992) 4583.}
\REF\mikov{A. Mikovi\'c, M. Ro\v cek, W. Siegel, P. Van Nieuwenhuizen,
           J. Yamron and A. E. Van de Ven, \pl\ {\bf B235} (1990) 106.}
\REF\mcjose{M. B. Green, C. Hull and J. L. V\'azquez-Bello,
               {\it Particles, Superparticles and Super-Yang-Mills},
               Preprint QMW-PH-92-08.}
\REF\Green{M. B. Green and C. M. Hull, \mpl\ {\bf A18} (1990) 1399.}
\REF\dirac {P. A. M. Dirac, \cjm\ {\bf 2} (1950) 129;
                            \pr {\bf 114} (1959) 924; in
              {\it Lectures on Quantum Mechanics}, ( Yeshiva
              University, 1964).}
\REF\vergara{M. Henneaux, C. Teitelboim and J. D. Vergara,
             {\it Gauge Invariance for Generally Covariant Systems},
              Preprint PAR-LPTHE-92-18/ULB-PMIF-92-03.}
\REF\barducci{A. Barducci, R. Casalbuoni, D. Dominici and R. Gatto,
                 {\it Unification of BRST, Paris-Sourlas and Space-time
                    Symmetries}, Preprint UGVA-DPT-1988-11-598.}
\REF\bello{J. M. L\'opez, M. A. Rodriguez, M. Socolovsky and
           J. L. V\'azquez-Bello, \rmf\ {\bf 34} (1988) 452.}
\REF\fredy{D. Z. Freedman and P. K. Townsend, \np\ {\bf B177} (1981) 282.}
%
%
\REF\west{A. Neveu and P. West, \pl\ {\bf B182} (1986) 343;
          W. Siegel and B. Zweibach, \np\ {\bf B282} (1987) 125;
          W. Siegel, \np\ {\bf B284} (1987) 632.}
\REF\kapasym{W. Siegel, \pl\ {\bf B128} (1983) 397.}
\REF\ghone{M. B. Green and C. M. Hull, {\it The Covariant Quantization of the
               Superparticle}, Presented at Texas A. and M. Mtg. on String
               Theory, College Station 1989.}
\REF\sieg{W. Siegel, \pl\ {\bf B203} (1988) 79.}
\REF\littleg{T. Kugo and P. Townsend, \np\ {\bf B221} (1983) 357.}
\REF\susybook{S. J. Gates Jr., M. T. Grisaru, M. Ro\v cek and W. Siegel,
                    {\it Superspace or One Thousand and One Lessons
                    in Supersymmetry } (Benjamin/Cummings, 1983).}
\REF\equis{L. Brink, M. B. Green and J. H. Schwarz,
                                  \np\ {\bf B223} (1983) 125.}
\REF\equistates{T. J. Allen, \mpl\ {\bf A2} (1987) 209.}
\REF\MandS{A. Mikovi\'c and W. Siegel, \pl\ {\bf B209} (1988) 47;
           R. Kallosh, A. Van Proeyen and W. Troost,
                                      \pl\ {\bf B212} (1988) 428.}
\REF\sokat {E. Sokatchev, \pl\ {\bf B169} (1986) 209;
                          \cqg\ {\bf 4} (1987) 237.}
\REF\bhtnew{L. Brink, M. Henneaux and C. Teitelboim,
                                \np\ {\bf B293} (1987) 505;
            E. R. Nissimov and S. J. Pacheva, \pl\ {\bf B189} (1987) 57.}
\REF\sspnews{M. Cederwall, Preprint ITP-G\"oteborg 89-15 (1989).}
%
%
%

\chapter {\it Introduction.}

All gauge theories, including super-Yang-Mills theories (SYM), are
characterized by a common feature, namely, unphysical degrees of freedom.
These unwanted degrees are contained within the fields
and must be fixed by an appropriate mechanism (gauge fixing condition)
when quantizing the model.
In the lagrangian formulation the gauge fixing condition is implemented by a
gauge breaking term which is added to the original gauge invariant classical
action so as to make the quantum action nondegenerate [\kugo ].
However, in general, a gauge fixing term might spoil other global symmetries
of the theory, like for example relativistic covariance.
The importance of covariant quantization is that the global
symmetries can be used to simplify calculations in the
quantum theory, and therefore it is recommended to leave these global
symmetries manifest.
%
%
Although, there have been
numerous contributions in this direction since the
work by Faddeev and Popov (FP) [\popov ], the Batalin and Vilkovisky (BV)
formulation encompasses all previous developments [\BV ].
Covariant quantization of
superparticle and superstring theories typically
requires the introduction of an infinite number of ghost fields
(infinite reducible gauge theories) [\book ], and can be approached
using the BV quantization procedure.

The Brink-Schwarz-Casalbouni (BSC) superparticle, which is a supersymmetric
extension of the standard relativistic particle [\casal ,\brink ],
has many properties in common with superstrings [\book ,\polyakov ,\romans ].
In particular, the D=10 superparticle describes
the dynamics of the zero-modes of the D=10 superstring
[\romans
-\huqone ], and so it is often used as a toy model.
The application of the BV procedure to the classical
BSC superparticle action has lead the reproduction of an incorrect physical
spectrum, which is known from a light-cone gauge (non-covariant)
quantization method [\delius ]. Although several modified superparticle
actions have been proposed with the intention of solving this
problem [\siegone
-\markov ], a truly covariant
quantization of the BSC model has not yet been found
[\polyakov ,\twisted ,\linds ].
However, in recent works [\huq
-\Green ], new proposed
superparticle models have the same physical spectrum as the original
BSC superparticle.

The purpose of this manuscript is to review some aspects of the various
superparticle models. It would take far to long to go into all the
details, however, these can be found in the original papers
[\casal ,\brink ,\siegone ,\kallosh ,\markov ,\huq ,\mikov ,\mcjose ].

\chapter {\it Local Reparametrization Invariance.}

Let us
consider a dynamical system formulated in a phase-space
with coordinates $(q_i ,p^i )$ $(i=1,\dots ,N)$ where $q_i$ are the
canonical coordinates and $p^i$ are their corresponding conjugate momenta.
The canonical Hamiltonian of the system is $H_0$. Let us also suppose
that there are \lq $m$' first-class constraints $G_a (q,p)$,
and they satisfy [\dirac ]
$$\{ G_a ,G_b \} = {f^c}_{ab} G_c\eqn\frisky$$
Here $\{ ,\}$ denotes Poisson brackets.
\footnote
\heartsuit{The Poisson bracket of two functions $f$ and $g$ in the phase-space
           with $N$ degrees of freedom is defined by
           $\{ f,g\} =\sum^{N}_{i=1} \Bigl[{{\grad f}\over {\grad q_i}}
                                               {{\grad g}\over {\grad p^i}}
                                         - {{\grad g}\over {\grad q_i}}
                                               {{\grad f}\over {\grad p^i}}
                                     \Bigr].$}
The constraints \frisky\ must
satisfy the following {\it stability} condition
$$\{ H_0 ,G_b \} = {V_b}^a G_a\;,\qquad (a,b =1,\dots ,m).\eqn\stably$$
The canonical
action is
$$S =\inttau \Bigl[ p^i {\dot q}_i -H_0 -\lambda^a G_a\Bigr]\eqn\canony$$
where
$\lambda^a (\tau )$ are Lagrange multipliers, and $\tau$
parametrizes the world-line in the phase-space.
The action \canony\ is made
invariant under a local gauge transformation, generated by the
constraints $G_a (q,p)$, by choosing convenient transformation
properties for the $\lambda$'s and appropriate boundary conditions
for the infinitesimal gauge parameters of the transformation
$\epsilon^a (\tau )$. The infinitesimal gauge transformations for \canony\
are given by [\vergara ,\barducci ]
$$\delta_\epsilon q_i =[q_i ,G_a \epsilon^a ],\qquad
  \delta_\epsilon p^i =[p^i ,G_a \epsilon^a ],\eqn\victim$$
and
$$\delta_\epsilon\lambda^a ={{\grad\epsilon^a}\over {\grad\tau}}
                            +[\epsilon^a ,H_0 +\lambda^b G_b ]
                            -{f^a}_{bc}\epsilon^b\lambda^c
                            -\epsilon^b {V_b}^a \;,\eqn\victimy$$
provided that
\victim\ and \victimy\ vanish at the endpoints of the world-line,
$\tau_1$ and $\tau_2$.
Precisely, the gauge parameters $\epsilon^a$ of the infinitesimal
transformations \victim\ and \victimy\
must satisfy the following {\it boundary} condition
$$\epsilon (\tau_1)=0,\qquad \epsilon (\tau_2)=0.\eqn\endpoints$$

The variation of the action \canony\ under the transformations
\victim\ and \victimy\ is proportional to the following
boundary term
$$\Bigl[ p^i {{\grad (\epsilon^a G_a )}\over {\grad p^i}}
           - \epsilon^a G_a \Bigr]{\Big\vert}^{\tau_2}_{\tau_1}
                                                        \;,\eqn\pointy$$
which cancels
as a result of \endpoints .
It has become customary to call a system with constraints
that are linear and homogeneous in the momenta {\it systems with
internal gauge symmetries}; Yang-Mills systems are of this type
[\vergara ].
For such systems
$$\Bigl[ p^i {{\grad G_a }\over {\grad p^i}} - G_a \Bigr] =0,\eqn\pointiux$$
vanishes
identically.
\footnote
\clubsuit{Although \pointiux\ is true for the Yang-Mills field,
          it does not hold, for example, for the
          Freedman-Townsend model [\fredy ].}

We can
choose a gauge in which $\lambda^a$ are constants,
simply by integrating \victimy . However, this is not a simple task
for local symmetries as one, in general, cannot integrate the
first-order differential equation \victimy\ with two simultaneous
boundary conditions.

\chapter {\it The Relativistic Scalar Particle.}

The
relativistic massive scalar particle is described by the evolution of a
massive point particle in a D dimensional space, and
its trajectory is represented by a
world-line ($x^\mu (\tau )$) where $\mu =0,1,\dots, D-1$ and $\tau$
parametrizes the world-line. The action for the relativistic scalar particle
is given by [\barducci ,\bello ,\west ]
$$S=-m\inttau\sqrt{-{\dot x}^\mu {\dot x}^\nu \eta_{\mu\nu}}
                                                     \;,\eqn\tivistic$$
where $\eta_{\mu\nu}$
is the D-dimensional Minkowski metric and
${\dot x}^\mu ={{\grad\xmu (\tau )}\over{\grad\tau}}$.
The action \tivistic\ can be
interpreted as the action of $D$ scalar fields in a $1$-dimensional
space-time, as it can be rewritten in the following form
$$S_{0} ={\half}\inttau [e^{-1}{\dot x}^\mu{\dot x}^\nu \eta_{\mu\nu}
                            - e m^2 ],\eqn\tivistix$$
where
$e$ is the {\it einbein} of the world-line. The canonical conjugate
momenta $\pmu$ is given by
$${{\grad L}\over {\grad {\xdot}^\mu}}=\pmu
               =m{ {{\xdot}_\mu}\over \sqrt{- {\xdot}^2} },\eqn\conjug$$
or
$$\pmu =e^{-1} \dot\xmu\eqn\conjugy$$
where
either \tivistic\ or \tivistix\ have been used, respectively,
as the Lagrangian
$L$ in the definition of the canonical conjugate momenta $\pmu$.
There is a first-class constraint
$$G_a =p^2 -m^2 \;.\eqn\choqui$$
{}From the reparametrization invariance of the relativistic action \tivistic ,
it follows that the canonical Hamiltonian vanishes identically,
$$H_0 = p_\mu \dot\xmu - L = 0.\eqn\vanity$$

The canonical action for the relativistic scalar particle is described in
the phase-space $(\xmu ,p_\mu )$ by
$$S_{canonical} =\inttau [p_\mu {\xdot}^\mu
                          -\half e (p^2 -m^2)],\eqn\scalarxy$$
where
$ e $ is a Lagrange multiplier. Therefore, the action \scalarxy\
can be made invariant under a local gauge transformation generated
by the constraint \choqui, by choosing convenient transformation properties
for $ e $ and appropriate boundary conditions for the infinitesimal
gauge parameters. Indeed,
the action \scalarxy\ is invariant under the following
$\tau$-reparametrization generated by the $G_a$
constraint
$$\delta_\epsilon \xmu =-2p^\mu \epsilon (\tau ),\qquad
  \delta_\epsilon e =\dot\epsilon (\tau ),\qquad
  \delta_\epsilon \pmu =0,\eqn\taudiff$$
where
$\epsilon (\tau )$ is the infinitesimal gauge parameter of the
transformation with fixed endpoints.
We also notice
that there is no problem with the massless limit in either
\tivistix\ or \scalarxy .

\chapter {\it BSC Superparticle.}

A superparticle is a particle moving in a superspace, and its
evolution is represented by a world-line in a $D$-dimensional
superspace $(\xmu (\tau ), \theta (\tau ))$ where $\mu =0,1,\dots ,D-1$;
$\theta$ is an anti-commuting Majorana-Weyl spinor and $\tau$
parametrizes the world-line.
The superparticle mechanics action proposed by Brink and Schwarz
is given in first-order form by
[\casal ,\brink ]
$$S_{BSC} =\inttau [p_\mu{\xdot}^\mu -i\theta\psh\dot\theta
                                        -\half e\;p^2 ],\eqn\bscaction$$
where
$\dot\theta ={{\grad\theta}\over {\grad\tau}}$, $\pmu$ is the momenta
of the superparticle and $e$ is the {\it einbein} on the world-line.
\footnote
\spadesuit{ It is often convenient to use a 16-component spinor notation
      to distinguish chirality, so that a right-handed Majorana-Weyl spinor
      $\psi_A$ has lower spinor index $(A=1,\dots ,16)$. In this notation,
      the supercoordinates has components
      $\thetaA$, $\theta\gammu\dot\theta=\thetaA (\gammu )^{AB}\dot\thetaB$,
      ${\psh}_{AB} =p_\mu (\gammu )_{AB}$, $(\gammu )_{AB} =(\gammu )_{BA}$,
      and so on.}
A variation of \bscaction\ with respect to $e$ implies the massless
Klein-Gordon equation $p^2 =0$. The remaining classical field
equations are
$$\dot\pmu =0,\qquad \psh\dot\theta =0,\qquad
  \pmu =e^{-1} [\dot\xmu -i\theta\gammu\dot\theta ].\eqn\extraequ$$

The action \bscaction\ is invariant under $\tau$-reparametrization
(generated by the $G_a =p^2$ constraint) together with
rigid space-time supersymmetry transformations
$$\delta_\epsilon \theta =\epsilon\;,\qquad
  \delta_\epsilon \xmu =i\epsilon\gammu\theta\;,\qquad
  \delta_\epsilon \pmu =\delta_\epsilon e =0,\eqn\rigid$$
where
the infinitesimal parameter $\epsilon^A$ is a constant Grassmann-valued
space-time anti-commuting spinor. The action \bscaction\ is also invariant
under a local fermionic symmetry [\kapasym ]
$$\eqalign{&\delta \theta =\psh\kappa ,\qquad
            \delta \xmu =-i\kappa\psh\gammu\theta ,\cr
           &\delta e =4i\dot\theta\kappa ,\qquad
            \delta \pmu =0 ,
            \cr}\eqn\fermisym$$
where
the infinitesimal gauge parameter $\kappa =\kappaA$ is a Majorana-Weyl
space-time spinor.
Although the superparticle action \bscaction\ is formulated in a superspace
with coordinates $(\xmu ,\thetaA )$, it lacks the conjugate momenta associated
with the supercoordinate $\thetaA$. Consequently, the canonical structure
of \canony\ is broken. In addition, the gauge algebra contains some extra
symmetries when the equations of motion are not used (off-shell).
In particular, the commutator of two fermionic symmetries gives a linear
combination of world-line diffeomorphisms plus a new transformation of the
form [\ghone]
$$\delta \xmu =p^2 v^\mu ,\qquad
  \delta e = -2\dot p_\mu v^\mu ,\eqn\extradiffs$$
where
$v^\mu$ is a bosonic vector parameter. This is a local symmetry of the
action \bscaction\ which is trivial, but it is needed to close
the gauge algebra off-shell.

\chapter {\it Siegel Superparticles.}

An alternative
action which restores the canonical form of \canony , should include a
conjugate momenta associated with the spinor coordinate $\thetaA$.
An action of this type was proposed by Siegel [\siegone ] for a
manifest space-time supersymmetry invariance by introducing
a gauge field $\psiA$ (referred as $SSP1$ superparticle
or {\cal AB} system).
The $SSP1$ action is [\siegone ,\siegel ]
$$S_{SSP1} =\inttau [p_\mu\dot\xmu +i\hat\theta\dot\theta
                     -\half e p^2 +i\psi\psh d ]\eqn\sspone$$
where
$d^A$ is a fermionic space-time anticommuting spinor introduced
so that the Grassmann coordinate $\thetaA$ has a conjugate momenta
${\hat\theta}^A =d^A -\psh^{AB}\thetaB$. The superparticle action \bscaction\
is obtained from \sspone\ by setting $d=0$.
A variation of \sspone\ with respect to $e$ and $\psiA$ implies,
respectively, the following $G_a$
constraints
$$G_e =p^2 ,\qquad G_\psi = \psh d,\eqn\gees$$
so that
the non-derivative terms in \sspone\ are the product of Lagrange
multipliers $\lambda^i$ with constraints $G_a$.
The remaining classical equations of motion
are
$$\eqalign{&\dot\pmu =0,\qquad \dot{\hat\theta} =0,\qquad
            \dot d =2\psh\dot\theta ,\cr
           &\dot\theta =\psh\psi ,\qquad
            \pmu=e^{-1} [\dot\xmu -i\theta\gammu\dot\theta +i\psi\gammu d].
            \cr}\eqn\remeqs$$

The algebra of constraints $G_a$ is
$$ \{ G_e ,G_e \}=0, \qquad \{ G_e ,G_\psi \} =0,
                  \qquad \{ G_\psi ,G_\psi \} =2 \psh G_e ,\eqn\algygee$$
subject to
the reducibility of the constraints. Indeed, an important feature of this
superparticle model is the reducibility of the constraints \gees , as
they are linearly
dependent
$$G_e\; d - \psh \; G_\psi = 0.\eqn\reducible$$
Therefore,
\reducible\ expresses that the superparticle \sspone\ is
a system with reducible constraints. In addition,
the $G_a$ are not only constraints, but the generators of a number of
local gauge symmetries by choosing convenient transformations
properties for the Lagrange multipliers and appropriate boundary
conditions for the infinitesimal gauge parameters of the transformations.
The action \sspone\ is then invariant under rigid Poincar\'e transformations
together with the rigid space-time supersymmetry and a number of
local symmetries
[\twisted ,\sieg ].
It is convenient to combine the reparametrization invariance with a
{\it trivial} symmetry
\footnote
\spadesuit{A trivial symmetry is one under which all fields transform
           into equations of motion. So that an action $S(\phi^i )$ dependent
           fields $\phi^i$ will automatically be invariant under local
           transformations of the form
           $\delta\phi^i =\lambda J^{ij} (\phi ) \delta S/\delta\phi^i $
           (with local parameter $\lambda$ ) provided that $J^{ij}$ is
           (graded) anti-symmetric [\twisted ].}
to obtain a local bosonic {\cal A} symmetry
$$\delta \xmu =p^\mu \xi ,\qquad \delta e =\dot\xi, \eqn\Asym$$
The local
fermionic symmetry (sometimes referred as {\cal B} symmetry) is given
by
$$\eqalign{&\delta \theta =\psh\kappa ,\qquad
            \delta \xmu =-i\kappa\psh\gammu\theta +id\gammu\kappa ,\cr
           \delta e =&4i\dot\theta\kappa ,\qquad
            \delta \psi =\dot\kappa ,\qquad
            \delta \pmu =0 , \qquad
            \delta d =2p^2 \kappa ,\cr}\eqn\Bsym$$
where
$\kappaA$ is an anticommuting Majorana-Weyl spinor. The action \sspone\
is also invariant under symmetries that act only on the gauge fields,
a generalized local bosonic symmetry and global space-time supersymmetries.
However, these further symmetries
are not needed to close the gauge algebra of contraints and we shall
ignore their status in the discussion given below, but its consequences
are considered elsewhere [\twisted ].

There is another reformulation to the $SSP1$ superparticle action which
includes a further gauge field $\chi$ so that one of the further global
space-time symmetries of the $SSP1$ action is turned into a local
symmetry [\markov ].
The $SSP2$ superparticle action (sometimes referred as the {\cal ABC} system )
is then an extention of the $SSP1$ action
by the addition of a bilinear term in the field $d^A$, and it
guarantees also the closure of the gauge algebra [\romans ,\markov ].
The $SSP2$ action is [\markov ,\sieg ]
$$S_{SSP2} =\inttau [p_\mu\dot\xmu +i\hat\theta\dot\theta
                     -\half e p^2 +i\psi\psh d
                     +\half d\chi d]\eqn\ssptwo$$
where
$\pmu$ is the momentum conjugate to the space-time coordinate $\xmu$, while
${\hat\theta}^A =d^A -\psh^{AB}\thetaB$ is conjugate to the Grassmann
super-coordinate $\thetaA$, and $e$, $\psiA$ and $\chi_{AB} =-\chi_{BA}$
are gauge fields which are also Lagrange multipliers imposing some
classical constraints.
The originial $S_{BSC}$ superparticle action \bscaction\
is given by setting $d=0$, while the $SSP1$ action is given by setting
$\chi =0$ in \ssptwo .

A variation of \ssptwo\ with respect to $e$, $\psi$ and $\chi$ implies,
respectively, the following classical
constraints
$$G_e =p^2 ,\qquad G_\psi =\psh d ,\qquad G_d =d\; d .\eqn\geestwo$$
These
constraints satisfy the algebra
$$\eqalign{&\{ G_\psi ,G_\psi \} =2\psh\; G_e ,\qquad
            \{ G_\psi ,G_d \} =2 d\; G_e ,\cr
           &\{ G_d , G_d \}_{ABCD} =4 (\psh )_{A[C}\; {G_d}_{D]B}
                                     - 4(\psh )_{B[C}\;{G_d}_{D]A}\;,\cr
           &\{ G_e ,G_e \} =0,\qquad
            \{ G_e ,G_\psi \} =0,\qquad
            \{ G_e ,G_d \} =0. \cr}\eqn\algetwo$$
However,
the right-hand sides of \algetwo\ are not unique
due to certain linear relations among the constraints \geestwo .
These are given by
$$\psh\; G_\psi - d\; G_e =0, \qquad
  \psh\; G_d -d\; G_\psi =0, \qquad
                d\; G_d =0,  \eqn\reductwo$$
and so on.
Therefore, they imply that the SSP2 superparticle model \ssptwo\ is also a
reducible system.
The remaining classical equations of motion are
$$\eqalign{&\dot\pmu =0,\qquad \dot{\hat\theta} =0,\qquad
            \dot d =2\psh\dot\theta ,\cr
           &\dot\theta =\psh\psi +i\chi d ,\qquad
            \pmu=e^{-1} [\dot\xmu -i\theta\gammu\dot\theta +i\psi\gammu d].
            \cr}\eqn\remyequs$$

The $SSP2$ action has a large number of symmetries which generalize
the ones found for the earlier superparticle models. It is invariant
under rigid Poincar\'e transformations together with the rigid space-time
supersymmetry. The action \ssptwo\ is also invariant under local gauge
transformations generated by the constraints \geestwo , which are referred
as the {\cal A}, {\cal B} and {\cal C} symmetries [\twisted ,\sieg ].
Explicitly, the local {\cal A} symmetry is given
by
$$\delta \xmu =\pmu\xi ,\qquad \delta e =\dot\xi, \eqn\Asymtwo$$
the {\cal B} symmetry is
$$\eqalign{&\delta \theta =\psh\kappa ,\qquad
            \delta \xmu =-i\kappa\psh\gammu\theta +id\gammu\kappa ,\cr
           \delta e =& 4i\dot\theta\kappa ,\qquad
            \delta \psi =\dot\kappa ,\qquad
            \delta \pmu =0 , \qquad
            \delta d =2p^2 \kappa ,\cr}\eqn\Bsymtwo$$
and
the {\cal C} symmetry is
$$\eqalign{&\delta \theta =-i\eta d ,\qquad
            \delta \xmu =-id\eta\gammu\theta ,\qquad
            \delta d= i\psh\eta d ,\cr
           &\delta e = -\psi\eta d ,\qquad
            \delta \chi =\dot\eta + i (\chi\psh\eta -\eta\psh\chi ) ,
            \cr}\eqn\Csymtwo$$
where
$\kappaA$ is a fermionic spinor parameter, while $\eta_{AB} =-\eta_{BA}$ is a
bosonic bispinor parameter associated with the gauge field
$\chi_{AB}$. There are also a number of local symmetries
that act only on the gauge fields and their presence reflects ambiguities
in the definition of the {\cal A}, {\cal B} and {\cal C} symmetries and
their relations among the constraints.

\chapter {\it Gauge Fixing and Quantization.}

We now
consider  both the counting of the fields and the choice of gauge conditions
that will fix the gauge invariances for the above superparticle models.
For the classical relativistic scalar particle \scalarxy , there are
10 degrees of freedom corresponding to the ten components of $\xmu$. The
degree of freedom corresponding to the field $e$ is gauged away, while
the momentum $\pmu$ is an auxiliary field and so can be eliminated by
its equation of motion.
In the quantum theory, the net number of degrees of freedom
is given by a {\it graded} total count. The momentum $\pmu$ is still
an auxiliary field, while the gauge invariance is fixed. There is now one
negative degree of freedom corresponding to the ghost of diffeomorphisms
giving a graded total of $10+1-1 =0$  (off-shell).

For the classical BSC superparticle \bscaction , the counting of degrees
of freedom for the bosonic sector is the same, as before.
For the fermionic sector, however, there are 16 degrees of freedom
corresponding to the components of the Majorana-Weyl spinor $\thetaA$.
The fermionic local symmetry allows the choice of a {\it non-covariant}
(physical) gauge in which half of the 16-components of the fermionic spinor
coordinate $\thetaA$ are gauged away, while the eight surviving components
of $\thetaA$ are self-conjugate, leaving a total net number of 8 fermionic
degrees of freedom.
In the quantum theory, the {\it graded} counting of degrees of freedom
for the bosonic sector still remains the same. For the fermionic sector,
there is a sequence of 16-component spinors
$(\theta ,\kappa_1 ,\kappa_2 ,\dots ,\kappa_n ,\dots )$  corresponding
to the fermionic spinor $\thetaA$ and the corresponding ghosts degrees
of freedom for the fermionic local symmetry $\kappa$, giving a graded
total of $16\times (1-1+1-1+\cdots )$ degrees of freedom (the alternating
sign corresponds to alternating statistics in the fields).
This ill-defined series is regularized to give a
total of $16\times\half =8$
degrees of freedom, which is in agreement with the classical
counting [\twisted ,\ghone ].

In a light-cone gauge (non-covariant), the reparametrization invariance
is used to set the gauge field $e$ to be a constant and the fermionic
symmetry is used to eliminate half of the 16-components of $\thetaA$.
An $SO(9,1)$ vector $\xmu$ $(\mu =0,1,\dots ,9)$ decomposes into an
$SO(8)$ vector $x^i$ $(i=1,\dots ,8)$ and two singlets $\xplus ,\xmin$
$(\xpm =x^0 \pm x^9 )$, so that if $\xplus$ is set equal to the solution of
its equation of motion, then $\xmin$ is determined by solving $p^2 =0$.
A 16-component spinor of $SO(9,1)$ decomposes into two 8-component $SO(8)$
spinors, so that $\gamma^+ \theta =0$ gauges away exactly eight of the
16-components of $\thetaA$. The remaining eight bosonic $x^i$ and eight
fermionic $\theta^a$ $(a=1,\dots ,8)$ variables transform as a vector and
spinor, respectively, of the little group $SO(8)$
[\book ,\twisted ,\littleg ].
One can further break the Lorentz covariance
of the superparticle from $SO(8)$ to $U(4)$, since the eight surviving
fermionic components are selfconjugate. So that in a canonical approach
they can be decomposed into four coordinates $\theta^a_c$ $(a=1,\dots ,4)$
and four momenta $\pi^a_c$. The physical content of the superparticle
is then given by the fields in the component expansion of a complex
superfield $\Phi (\pmu ,\theta^a_c )$ which satisfies the constraint
$p^2 \Phi (\pmu ,\theta^a_c )=0$ and a reality condition.
The component expansion gives $1+6+1 =8$
real bosonic components, and $4+4 =8$ real fermionic components.
Such physical content of the BSC superparticle fit together to give
the spectrum for the $N=1$ super-Yang-Mills Theory [\susybook ,\equis ].

For the classical SSP1 superparticle \sspone , there are 10 degrees of
freedom corresponding to $\xmu$, the field $e$ is gauged away and $\pmu$
is eliminated by its equation of motion.
There are also 16 degrees of freedom corresponding to $\thetaA$.
However, the fermionic symmetry can be used to gauge away eight of the
16 components of $\theta$ which is in accord with the expected physical
spectrum of the theory.
In a {\it covariant quantization} it is necessary to find a
covariant gauge choice for both the reparametrization and fermionic
symmetries.
The reparametrization invariance can be fixed by imposing a constraint
on the einbein $e=constant$, while the fermionic invariance can be fixed
by imposing a gauge condition on the fermionic gauge field, $\psi =0$.
However, a light-cone gauge quantization of the SSP1 reveals that its
corresponding spectrum contains $2^8$ states and it is therefore
not equivalent to the $N=1$ $BSC$ superparticle action \bscaction\ which
has a spectrum of $2^4$ states [\twisted ,\equistates ].
It was emphasized previously that the SSP1 model is invariant under
a further global space-time supersymmetry so that the corresponding
superalgebra of the generators of these global supersymmetries have a
{\it twisted} nature.
\footnote
\clubsuit{If the supercharges generating the global space-time
          supersymmetries are denoted by $Q_1$ and $Q_2$, respectively, the
          corresponding superalgebra have the twisted algebra
          $\{ Q^A_1 ,Q^B_1 \} =-\{ Q^A_2 ,Q^B_2 \} =2(\gammu )^{AB}\pmu$,
          and $\{ Q^A_1 ,Q^B_2 \} =0$. The relative minus sign means
          a $N=2$ twisted supersymmetry which has automorphism group
          $SO(1,1)$, and it also leads the presence of negative norm
          states [\twisted ].}
Therefore,
the $N=1$ SSP1 superparticle is on-shell equivalent to a
{\it twisted} version of the $N=2$ SSP0 superparticle that has negative
norm states [\twisted ].

The SSP2 superparticle \ssptwo\ is an extention of the SSP1 action by the
addition of a bosonic bi-spinor $\chi_{AB}$ which acts as a Lagrange
multiplier imposing a constraint that is intended to remove those unwanted
negative norm states of the SSP1.
The extra term breaks the further global space-time
supersymmetry of the SSP1, so that the SSP2 action is left invariant under a
single global space-time supersymmetry.
In a light-cone gauge analisis of the SSP2,
it is found that there are again eight bosonic degrees of freedom and eight
$\theta$'s {\it plus} eight independent $d$'s, but also there is a residual
$SO(8)$ symmetry which is used to gauge $d$ to a constant, so that the
physical spectrum of the SSP2 consists of eight bosons and eight fermions,
corresponding to the spectrum for the $N=1$ super-Yang-Mills
theory [\twisted ,\equis ,\MandS ].
However, covariant quantization of the SSP2 superparticle yields
unsatisfactory results, because its BRST operator does not give the correct
BRST cohomology classes [\linds ].

To summarize, two reformulations of the original superparticle action
BSC \sspone\ have been proposed which include a gauge field for the
local fermionic symmetry. Neither of these reformulated models give
satisfactory results, but their quantization illustrates a number of
interesting features. Covariant quantization of any of these superparticles
or further modifications require an infinite number of ghost fields which
reflects the ambiguities on the infinite reducibility of the constraints.

%

\chapter {\it Quadratic Superparticles.}

Covariant
quantization of the first-class {\cal A B C}
system [\linds ], either by BRST or BV methods has failed, because the BRST
operator does not give the correct cohomology [\mikov ].
In Ref. [\mikov ], it was shown that the BRST quantization of any set of
constraints forming a compact gauge algebra should contain a singlet
of the gauge group.
The above problem for the {\cal A B C} system can be avoided
by the appropriate modification of the constraints. In Ref. [\mikov ], two
formulations were also considered for solving the difficulties of the
{\cal A B C} system
(referred as {\it first} or {\it second-ilk} superparticles).
The {\it first-ilk} or {\cal A B C D} superparticle is defined by
introducing a new fermionic variable $\Gamma_a$ (which satisifes
$\{ \Gamma_a, \Gamma_b \} =2\eta_{ab}$) and a new constraint {\cal D}.
The full set of constraints for this system is given by
[\unoilk ,\covilk ]
$$\eqalign{& {\cal A} =p^2 ,\qquad
             {\cal B} = \psh d ,\qquad
             {\cal D} = p\cdot\Gamma ,\cr
   {\cal C}_{\alpha\beta} =&\third (\gamma^{abc})_{\alpha\beta}C_{abc}
          = d_{[\alpha} d_{\beta]}
              +\half (\gamma^{abc})_{\alpha\beta}\;p_a\Gamma_b\Gamma_c
.\cr}\eqn\abcdgees$$
The
constraints {\cal A}, {\cal B}, {\cal C} and {\cal D} satisfy the
algebra [\unoilk ,\covilk ]
$$\eqalign{\{ {\cal B},{\cal B}\} =&2\psh  {\cal A},\qquad
            \{ {\cal C},{\cal B}\}_{\alpha\beta}^{\ \ \ \gamma}
                                   =-4\delta_{[\alpha}^{\ \ \gamma}
                                       d_{\beta]} {\cal A},\qquad
            \{ {\cal D}, {\cal D} \} = 2{\cal A}  \;,\cr
   \{ {\cal C}, {\cal C}\}_{\alpha\beta\gamma\delta}
           &=4(\psh )_{\alpha [\gamma} {\cal C}_{\delta ]\beta}
                 -4 (\psh )_{\beta [\gamma} {\cal C}_{\delta ]\alpha} \cr
           &\qquad          -\half (\gamma_{cd}^{\ \ a})_{\alpha\beta}
                        (\gamma^{bcd})_{\gamma\delta}
                       [\half \Gamma_{[a} \Gamma_{b]} {\cal A}
                                     + p_{[a}\Gamma_{b]} {\cal D}].
\cr}\eqn\abcdalgy$$
However,
the right-hand sides of \abcdalgy\ are not unique due to the reducibility
of the constraints {\cal A}, {\cal B}, {\cal C} and {\cal D}.
The system
{\cal A B C D} is infinitely reducible in the
sense of Batalin and Vilkovisky [\BV ].

On the other hand, the {\it second-ilk} superparticle consists
of an infinite number of constraints and an infinite number of
spinorial coordinates and momenta.
The full set of constraints for
this model is given by [\mikov ]
$${\cal A} =p^2 ,\qquad
  {\cal B} =\psh q_0 ,\qquad
  {\cal C}_n =d_{2n} +q_{2n+2}\;,\eqn\secdilk$$
where
$d_n =-i\hat\theta_n +\psh\theta_n$ and $q_n =-i\hat\theta_n -\psh\theta_n$
$(n=0,1,\dots )$ are fermionic space-time anticommuting spinors,
$\hat\theta_n$ are conjugate momenta to the Grassmann supercoordinates
$\theta_n$ $(n=0,1,\dots )$, and satisfy
$\{ \hat\theta_m ,\theta_n \} =i\delta_{mn}$.
The constraints \secdilk\ are also infinitely reducible.

In Ref. [\mikov ], a new classical action was found based upon
the constraints \secdilk . This {\it second-ilk} superparticle describes
and infinite sequence of $SSP1$ superparticles plus a term that breaks
down all the twisted supersymmetries of the infinite sequences of
$SSP1$ superparticles.
This model is formulated in a ten-dimensional
superspace with coordinates $(\xmu ,\theta_0 ,\dots ,\theta_{2n} ,\dots )$,
where $(\theta_0 ,\dots ,\theta_{2n} ,\dots )$ are anti-commuting spinors.
The action is [\belli ,\mikov ]
$$\eqalign{S_{sec-ilk} =\inttau\Bigl[ p_\mu\dot\xmu -\half e p^2
                        &+\sum^{+\infty}_{n=0}\dot\theta_{2n}\hat\theta_{2n}
                           -\psi_1 \psh q_0 \cr
                        &-\sum^{+\infty}_{n=0}\lambda_{2n+1}
                                  (d_{2n} + q_{2n+2})\Bigr],\cr}
\eqn\ilkaction$$
where
$\hat\theta_{2n} =d_{2n} -\psh \theta_{2n}$ are the conjugate momenta to
the supercoordinates $\theta_{2n}$, and $e$, $\psi_1$ and $\lambda_{2n+1}$
are gauge fields which are also Lagrange multipliers imposing the infinite
set of classical constraints {\cal A}, {\cal B} and ${\cal C}_n$.

The remainig classical field equations of motion are [\belli ]
$$\eqalign{ \dot\pmu =0,\qquad
           &\psh\dot\theta_0 -p^2\psi_1 =0,\qquad
            \dot\theta_{2n}\psh-i\lambda_{2n-1}\psh =0 ,\cr
            \pmu=(2e)^{-1} [\dot\xmu -&i\psi_1\gammu d_0
                            +4i\psi_1 \gammu\theta_0
                            -i\sum^{+\infty}_{n=0}
                                      \dot\theta_{2n}\gammu\theta_{2n} \cr
           &\hskip 3.0cm   +2\sum^{+\infty}_{n=0}
                                      \lambda_{2n+1}\gammu\theta_{2n+2}].
            \cr}\eqn\ilkremy$$

The {\it second-ilk} superparticle \ilkaction\ is invariant under
a number of local gauge symmetries generated by the infinite
set of constraints {\cal A}, {\cal B} and ${\cal C}_n$ [\belli ].
It is also invariant under local
gauge symmetries that act only on the gauge fields and their presence
reflects ambiguities on the infinite reduciblity of the constraints
{\cal A}, {\cal B} and ${\cal C}_n$.

There is another similar superparticle action of the {\it second-ilk} type.
It was found when a special combination of a certain $N=2$ supersymmetry
was promoted to a local symmetry, and whose presence is crucial for obtaining
the correct BRST operator [\kallosh ,\huq ].
It starts by considering a $SSP1$ action and introducing an infinite
tower of anti-commuting fermionic variables, which are identified with
the zero ghost number sector of the pyramid of ghosts appearing in the
covariant quantization of the BSC superparticle.
A further term that breaks down the infinite tower of twisted $N=2$
supersymmetries is added and as a result the final action is invariant
under a global $N=1$ supersymmetry.
The action is [\huq ]
$$\eqalign{S_{sec-ilk} =\inttau\Bigl\{ p_\mu\dot\xmu -\half e p^2
                          &+\sum^{+\infty}_{n=0}\lambda_n
                                  (\dot\theta_n -\psh\psi_n )\cr
                          &-\sum^{+\infty}_{n=0}
                                 [(\lambda_n +i\psh\theta_n )
                                 +(\lambda_{n+1}-i\psh\theta_{n+1})]\chi_n
\Bigr\},\cr}\eqn\huqilky$$
where
$\psi_n$ is the gauge field for the fermionic symmetry at the {\it n'th}
level, while $\lambda_n$ is the conjugate momenta to the supercoordinate
$\theta_n$, and $e$ and $\chi_n$ are gauge fields which are also
Lagrange multipliers imposing the following infinite set of classical
constraints
$${\cal A}=p^2 ,\qquad
  {\cal B'}_n =\psh\psi_n \;,\qquad
  {\cal C'}_n = (\lambda_n +i\psh\theta_n )
               +(\lambda_{n+1} -i\psh\theta_{n+1} ).\eqn\huqgees$$

The superparticle action \huqilky\ is invariant under the usual {\cal A}
symmetry, together with an infinite sequence of fermionic symmetries
with infinitesimal anticommuting spinor parameters $\kappa_n$ and $\xi_n$,
respectively [\huq ].
There are also symmetries of the second-type which act only
on the gauge fields and reflect the infinite reducibility of the system.
\footnote
\spadesuit{A complete cohomology analysis of the BRST operator for the
           first-ilk superparticle \abcdgees\ was given in Refs.
           [\unoilk ,\covilk ], while similar cohomology analysis for the
           second-ilk superparticles \ilkaction\ and \huqilky\ were given
           in Refs. [\mikov ] and [\kallosh ,\huq ], respectively.
           These analysis showed that the cohomology reproduces the desired
           spectrum of the ten dimensional super-Yang-Mills theory.}

\chapter {\it Superparticles in Extended Spaces.}

Let us consider
further modifications of the superparticle that lead to free
BRST invariant quantum actions.  These new superparticle theories
are obtained by the addition of extra coordinates to the superspace
$(\xmu ,\thetaA )$, and their physical states are described either
by a superspace spinor or vector wave function satisfying some
linear or quadratic constraints [\equis ].
The wave function for these new models is a superfield whose physical
components are those of the super-Yang-Mills theory
[\mcjose ,\equis ].
In this section, we present a summary of these new superparticle
theories. A more complete treatment is presented in Ref. [\mcjose ].

We first seek superparticle theories formulated in an extended
superspace with coordinates $(\xmu ,\thetaA ,\phiA )$, where
$\thetaA$ and $\phiA$ are anti-commuting Majorana-Weyl spinors.
Here $(\xmu ,\thetaA )$ are the usual coordinates of the ten-dimensional
$N=1$ superspace and $\phiA$ is a new spinor coordinate.
In [\Green ] an extra
spinor coordinate $\phiA$ was introduced, together with
its conjugate momentum $\hatphiA$ and a momentum ${\hat\theta}^A$
conjugate to $\thetaA$. It is convenient to define
${\hat\theta}^A =d^A -{\psh}^{AB}\thetaB$.
The action is the sum of a free action
\footnote
\spadesuit{We suppress spinor indices and use a notation, so that
            $d\dot\theta =d^A {\dot\theta}_A$ ,
  $\theta\Gammu{\dot\theta} =\thetaA{(\Gammu )}^{AB}{\dot\theta}_B$,
  $d\chi d =d^A {\chi_{AB}} d^B$, $\hat\phi\Gammu\chi\Gamma_\mu\psh\phi =
  {\hat\phi}_A{(\Gammu )}^{AB}\chi_{BC}{(\Gamma_\mu )}^{CD}{\psh}_{DE}\phi^E$,
   etc.}
$$S_0 =\inttau \Bigl[p_\mu {\dot x}^\mu + i\hat\theta\dot\theta
               +i\hat\phi\dot\phi \Bigr],\eqn\zeroaction$$
plus the term
$$S_1 =\inttau \Bigl[-\half e p^2 + i\psi\psh d  + i\Lambda\psh\hatphi
                     + \half\hat\phi\Upsilon\hat\phi +\half d\chi d
                     +2 \hat\phi\Gammu\chi\Gamma_\mu \psh\phi
                                          \Bigr],\eqn\oneaction$$
where
$e$, $\psi^A$, $\chi_{AB}=-\chi_{BA}$, $\Lambda_A$ and
$\Upsilon^{AB} =-\Upsilon^{BA}$ are gauge fields
which are also Lagrange multipliers imposing some classical constraints.
A variation of \zeroaction\ and \oneaction\ with respect to $e$, $\psi$,
$\Lambda$, $\Upsilon$ and $\chi$ implies the following
classical constraints
$$\eqalign{& {\cal A}=p^2 \;, \qquad
             {\cal B}=\psh d \;, \qquad
             {\cal D}=\psh\hat\phi  \;,\cr
             {\cal G}=&{\hat\phi}_A{\hat\phi}_B \;, \qquad
             {\cal C}=d^A d^B -8(\hat\phi \Gammu )^{[A}
                                         (\Gamma_\mu \psh\phi )^{B]}.
\cr}\eqn\conditions$$
This new superparticle
action formulated in an extended space
has a large number of local gauge symmetries which
generalize the ones found for earlier models.
The covariant quantization
of this superparticle model was discussed in [\Green ] by
choosing the gauge $e=1$, with the other gauge fields set to zero.
Covariant quantization uses the BV procedure, as the gauge algebra
only closes on-shell, and requires an infinite number of ghosts fields.

There is another reformulation which leads to a spinor
wave function satisfying certain linear contraint [\equis ].
The superparticle action is [\mcjose ]
$$S_{spinor} =\inttau \Bigl[p_\mu \dot x^\mu +i\hat\theta\dot\theta
                        +i\hat\phi \dot\phi \Bigr],\eqn\sspthree$$
plus the term
$$\eqalign{S_{extra} =\inttau\Bigl[-\half ep^2  + i\psi\psh d
                                   +i\varphi\psh\hatphi
                                 & +i{\Lamnrs}\;d\;{\Gamnrs}\hat\phi \cr
                      &\qquad      -i\beta (\phi\hat\phi -1 )
                                   +\half\hat\phi\omega\hat\phi \Bigr],
\cr}\eqn\xtraspin$$
where,
as usual, $p_\mu$ is the momentum conjugate to the space-time coordinate
$x^\mu$, $d^A$ is a spinor introduced so that the Grassmann coordinate
$\theta$ has a conjugate momentum ${\hat\theta}^A =d^A -\psh^{AB}\theta_B$,
$\phi^A$ is also a new spinor coordinate and $\hat\phi_A$ its conjugate
momentum. The fields $e$, $\psi^A$, $\varphi_A$, $\Lamnrs$, $\beta$ and
${\omega}^{AB}$ are all gauge fields which
are also Lagrange multipliers imposing some finite set of classical
constraints.
These are given by
$$\eqalign{& {\cal A}=p^2 , \qquad
             {\cal B}=\psh d , \qquad
             {\cal D}=\psh\hat\phi =0 ,\cr
             {\cal G}={\hat\phi}_A{\hat\phi}_B ,&\qquad
             {\cal H}={\phiA}{\hat\phi}_A - 1 ,\qquad
             {\cal C}={\dA} ({\Gamnrs} {)^B}_A {\hat\phi}_B .\cr}
\eqn\newthreecon$$
The constraints \newthreecon\ are also infinitely reducible in the
sense of Batalin and Vilkovisky [\BV ].

In
the remainder of this section,
we are concerned with superparticle theories which lead to a
vector wavefunction satisfying either a linear or a quadratic constraint
[\equis ].
We first seek a superparticle action formulated in an extended
superspace with
coordinates $(x^\mu ,\theta_A ,\phimu )$ where $\theta_A$ is an anti-commuting
Majorana-Weyl spinor, and $\phimu$ is a vector field. Here
$(x^\mu ,\theta_A )$ are the coordinates of the usual 10-dimensional
N=1 superspace and $\phimu$ is a new vector coordinate.
The superparticle action (SSP-vector)
is given by [\mcjose ]
$$S_{0} =\inttau \Bigl[p_\mu {\dot x}^\mu + i\hat\theta\dot\theta
                       +{\hatphi}_\mu {\dot\phi}^\mu \Bigr],\eqn\newvecci$$
plus the term
$$\eqalign{S_{1}=\inttau \Bigl[-\half ep^2 + i\psi\psh d + \half d\chi d
                             & +2i{\hatphi}_\mu\chi{{\psh}^\mu}_\nu\phinu\cr
                             & -\omega \pmu {\hatphi}_\mu
                               +\half {\hatphi}_\mu\;\rhomunu{\hatphi}_\nu
                              \Bigr],
\cr}\eqn\xtravecci$$
where
$p_\mu$ is the momentum conjugate to the space-time coordinate
$x^\mu$, $\dA$ is a spinor introduced so that the Grassmann coordinate
$\theta$ has a conjugate momentum ${\hat\theta}^A =\dA -\psh^{AB}\thetaB$.
Here, $\phimu$ is a new vector coordinate and $\hatphi_\mu$ its conjugate
momentum.
The fields $e$, $\psi^A$, $\chi_{AB} =-\chi_{BA}\;$,
$\rho_{\mu\nu} =\rho_{\nu\mu}$ and $\omega$ are all Lagrange multipliers
imposing the following finite set of classical
constraints
$$\eqalign{& {\cal A}=p^2 , \qquad
             {\cal B}=\psh d , \qquad
             {\cal D}=p^\mu {\hatphi}_\mu , \cr
            {\cal G}&={\hatphi}_\mu {\hatphi}_\nu ,\qquad
             {\cal C}=\dA\dB +4{\hatphi}_\mu ({\psh^\mu}_\nu )^{AB}\phinu .
\cr}\eqn\verdevec$$
The {\cal A}, {\cal B}, {\cal C}, {\cal D} and {\cal G}
constraints are the generators of a number
of local gauge symmetries.
There are also symmetries of the second-kind which reflect the
infinite reducibility of the constraints {\cal A}, {\cal B}, {\cal C},
{\cal D} and {\cal G}.

We seek now a superparticle theory which lead
to a vector wave function satisfying a linear constraint [\equis ].
The superparticle action is formulated in an extended superspace with
coordinates $(\xmu ,\thetaA ,\phimu )$ where $\thetaA$ is an anti-commuting
Majorana-Weyl spinor. Here, $(\xmu ,\thetaA )$ are the coordinates of the
usual ten dimensional $N=1$ superspace and $\phimu$ is a new vector
coordinate. The new superparticle action (SSP-vector) is
given by [\mcjose ]
$$S_{vector}=\inttau \Bigl[ p_\mu {\dot x}^\mu
                              +i{\hat\theta}{\dot\theta}
                              +\hatphi_\mu {\dot\phi}^\mu
\Bigr],\eqn\azulzero$$
plus the term
$$\eqalign{S_{extra} =\inttau \Bigl[-\half e p^2  + i\phi\psh d
                              +\lambda \pmu {\hat\phi}_\mu
                         & +\half \hatphi_\mu\omega^{\mu\nu}\hatphi_\nu \cr
                         & +\beta (\phimu \hat\phi_\mu -1)
                              +\Upsimupha \Cmupha \Bigr],
\cr}\eqn\azulxtra$$
where
$\Cmupha = \hatphi_\mu \dA
           - \seventh \hatphi_\nu ({ {\Gamma_\mu}^\nu }{)^A}_B \dB$,
as usual, $\pmu$ is the momentum conjugate
to the space-time coordinate $\xmu$, $\dA$ is a spinor introduced so that
the Grassmann coordinate $\thetaA$ has a conjugate momentum ${\hat\theta}_A$,
and $\phimu$ is a new vector coordinate together with its conjugate
momentum ${\hat\phi}_\mu$.
The fields $e$ ,$\psiA$, $\lambda$,
$\omega^{\mu\nu} =\omega^{\nu\mu}$, $\beta$ and $\Upsimupha$ are all
Lagrange multipliers imposing some classical
constraints.
A variation of \azulzero\ and \azulxtra\ with
respect to $e$, $\phi$, $\lambda$, $\omega$
$\beta$ and $\Upsilon$ implies the following set of classical
constraints
$$\eqalign{&{\cal A}=p^2 ,\qquad
            {\cal B}=\psh d ,\qquad
            {\cal D}=\pmu\hatphi_\mu ,\cr
            {\cal G}=&\hatphi_\mu \hatphi_\nu ,\qquad
            {\cal C}={C_\mu}^A ,\qquad
            {\cal H}=\phi^\mu \hatphi_\mu -1,\cr}
\eqn\azulconstr$$
which are also
the generators of a number
of local gauge symmetries, together with symmetries of the second-kind
due to the reducibility of the constraints {\cal A}, {\cal B}, {\cal C},
{\cal D}, {\cal G} and {\cal H}.

Covariant quantization of any of the previous
superparticle extended models require the use of the Batalin
and Vilkovisky procedure, as the gauge algebra of their
constraints only closes on-shell, and calls for an infinite
number of ghost fields.

Finally, there are still other models for the superparticle
as the Sokatchev's harmonic superparticle [\sokat ], or
actions with light cone directions chosen in a covariant way using
dynamical variables [\bhtnew ]. However, the structure of these actions
is different from those we are considering here and shall not be
reviewed, but further details can be found elsewhere
[\polyakov ,\sokat ,\bhtnew ,\sspnews ].

\ack {The author wishes to thank C. M. Hull for
      reading of the manuscript and stimulating
      discussions. It is also acknowledge the help
      of C. S. Burrows for correcting errors present
      in an earlier draft.}

\vfil\eject
\refout

\end
\bye